\documentclass[12pt]{iopart}

\usepackage{iopams}
\usepackage{amsthm}
\usepackage{amssymb}
\usepackage{graphicx} 
\graphicspath{ {images/} }
\begin{document}

\title{Goodness of Isospin in Neutron Rich Systems from the Fission Fragment Distribution}


\author{Swati Garg and Ashok Kumar Jain}
\address{Department of Physics, Indian Institute of Technology Roorkee, Roorkee-247667, India}

\ead{swat90.garg@gmail.com}
\vspace{10pt}
\begin{indented}
\item[]{\today}
\end{indented}

\begin{abstract}
We present the results of our calculations for the relative yields of neutron-rich fission fragments emitted in $^{208}$Pb ($^{18}$O, fission) reaction by using the concept of the conservation of isospin and compare with the experimental data. We take into account a range of isospin values allowed by the isospin algebra and assume that the fission fragments are formed in Isobaric Analog States. We also take into account the neutron multiplicity data for various neutron-emission channels in each partition, and use them to obtain the weight factors in calculating the yields. We then calculate the relative yields of the fission fragments. Our calculated results are able to reproduce the experimental trends reasonably well. This is the first direct evidence of the isospin conservation in neutron-rich systems and may prove a very useful tool in their studies.
\end{abstract}


%

\section{Introduction}
Isospin was first introduced by Heisenberg in 1932~\cite{heisenberg}. It is a fundamental quantum number introduced to distinguish between two states of nucleon, the proton and the neutron. Then Cassen et al.~\cite{cassen} expanded the scope of Pauli principle by proposing the Generalized Pauli principle which includes the isospin degree of freedom in the wave function. As a result, the wave function should be antisymmetric under space, spin and isospin for Fermions. Generally, isospin remains a good quantum number in lighter nuclei as the Coulomb effects are small. Later, experimental findings by Anderson et al.~\cite{anderson} and Fox et al.~\cite{fox} concluded that isospin can be a useful quantity in heavier nuclei also.
The concept of isospin has been widely used in lighter $N \approx Z$ nuclei where it is a reasonably good quantum number~\cite{warner, satula}. It is, however, interesting to note that the purity of isospin may be restored in neutron-rich systems, which are now becoming more accessible in labs. The isospin impurity comes into picture mainly because of the Coulomb interaction which leads to the admixture of  $T=T_3$ with $T'=T_3$ +1 states~\cite{robson, robson1}. In 1962, Lane and Soper~\cite{lane} provided a theoretical basis to the observation that isospin can be useful in heavier nuclei and showed that the isospin impurity decreases with the neutron excess by a factor of 2/($N-Z$+2).  
The neutron-rich fragments emitted in nuclear fission of heavy nuclei appear to be a good testing ground for the goodness of isospin in neutron-rich systems. Jain et al.~\cite{jain} proposed the idea that the major features of neutron-rich fission fragments could be obtained by using the concept of isospin conservation. However, the ambiguity in assigning the isospin to fission fragments presented difficulties. In this short paper, we briefly present a scheme to assign isospin values to fission fragments and calculate the fission fragment distribution.
We calculate the relative yields of neutron-rich fission fragments emitted in HI fusion fission reaction $^{208}$Pb ($^{18}$O, f) and compare our results with experimental data from Bogachev et al.~\cite{bogachev} and Banerjee et al.~\cite{banerjee}. We use the isospin conservation to fix the value of isospin of the Compound nucleus ($CN$) and the two fragments, $F_1$ and $F_2$. Using the isospin part of the wave-function only, we calculate the relative intensities of the various fragments. We also incorporate the neutron multiplicity data from Bogachev et al.~\cite{bogachev} to include the weight factors of various $n$-emission channels. We find that our theoretical results are able to explain the relative yields of fission fragments in reasonable way.

\section{Formalism}
\label{sec:1}
Isospin algebra is similar to that of spin and satisfies the SU(2) group. For both neutrons and protons, isospin is the same i.e. $T$ = 1/2 but the third component of isospin $T_3$ is different, i.e. +1/2 for neutron and $–$1/2 for proton. For a ($N, Z$) nucleus,

\begin{equation}
T_3 = (N-Z)/2 
\end{equation}
and 

\begin{equation}
\mid(N-Z)/2\mid \leq T \leq (N+Z)/2
\end{equation}
We divide our formalism in two main parts. In the first part, we assign isospin to the $CN$ and the fission fragments and in the second part, we calculate the relative yields of fragments which are proportional to the square of the respective Clebsch-Gordan coefficients ($CGC$) arising in the isospin part of the wavefunction.

First, we consider a HI induced fusion-fission reaction leading to the formation of a $CN$ which further breaks into two fragments $F_1$ and $F_2$ with the emission of $n$ number of neutrons,
\begin{equation}
Y(T_Y,T_{3Y})+X(T_X,T_{3X}) \rightarrow CN(T_{CN},T_{3CN})
\rightarrow F_1(T_{F1},T_{3F1})+F_2(T_{F2},T_{3F2})
\end{equation}
where $T_Y, T_X, T_{CN}, T_{F1}$ and $T_{F2}$ are the total isospin values of projectile, target, $CN$, and the two fragments, respectively. From isospin conservation, the isospin of $CN$ should lie in the range $\mid T_{X}-T_Y \mid \leq T_{CN} \leq (T_{X}+T_{Y})$ and its third component is given as $T_{3CN}=T_{3Y}+T_{3X}$. We assume the target and projectile to be in the ground state. For the ground states of all nuclei considered in this work (because they are even-even), the total isospin quantum number is equal to its third component. So, it gives $T_Y=T_{3Y}$ and $T_X =T_{3X}$, and $ \mid T_{3X}-T_{3Y}\mid \leq T_{CN} \leq (T_{3X} +T_{3Y})$. But, from the isospin algebra, $T_{CN}\geq T_{3CN}$ and $T_{3CN}=T_{3Y}+T_{3X}$. It gives us a unique value of $T_{CN}$  i.e. $T_{CN}=T_{3CN}=T_{3Y}+T_{3X}$. For example, in the reaction under consideration $^{208}$Pb ($^{18}$O, f), $T_X$($^{208}$Pb) = $T_{3X}$($^{208}$Pb) = 22 and $T_Y$($^{18}$O) = $T_{3Y}$($^{18}$O) = 1. Thus isospin of CN can have three possible values,  $T_{CN}$ = 22$-$1, 22, 22+1 = 21, 22, 23. But $T_{CN}$ = 22+1 = 23, which leads to only one value of $T_{CN}$ = 23.

The $CN$ then breaks into two fragments with the emission of neutrons. Here, we introduce an auxiliary concept of residual compound nucleus ($RCN$) which is assumed to be formed after the emission of $n$ number of neutrons to simplify our problem. We do not consider the pre and post scission neutrons separately because the time difference between the two is very small, approximately equal to 10$^{-19}$ sec.

Now the main difficulty is to assign the total isospin to the fission fragments and $RCN$. For this, we invoke two conjectures of Kelson~\cite{kelson}. First one assumes that the neutron emission favors the formation of excited states with $T>T_3$. So we assign the isospin of $RCN$ to be $T_{RCN}=T_{F1}+T_{F2}$, which is its maximum possible value. Also, from the conservation of isospin, $T_{RCN}$ should lie in the range,
\begin{equation}
\mid T_{CN}-n/2 \mid \geq T_{RCN} \geq (T_{CN}+n/2)
\end{equation}

To assign isospin to various fission fragments, we use Kelson's second conjecture which suggests that the ``tendency to overpopulate highly excited states with $T >T_3$ in the primary fission products, carries largely over to the conventionally referred to Isobaric Analog states (IAS) in the observed products''. Thus, observed fission fragments should preferably form in IAS. This helps us in assigning total isospin values to the individual fission fragments. We choose three isobars corresponding to each mass number. Then we assign to $T_{Fi}$ the maximum value of $T_{3Fi}$ among the three isobars for that particular mass number because this will be the minimum value of $T$ required to generate all the members of any complete isobaric spin multiplet. For example, we have three isosbars of $A=100$, $^{100}$Ru, $^{100}$Mo, $^{100}$Zr having $T_3$ values as 6, 8, 10 respectively. So, we assign total isospin $T=10$ for the three isobars having $A=100$. We obtain in this way, the total isospin values of the $CN$, $RCN$ and the two fragments.

We now move to second part of the calculations that is to calculate the relative yields of fission fragments. We consider only the isospin part of the total wavefunction and make all the possible pairs of fragments for a particular $n$-emission channel. For a particular pair of fragments in $n^{th}$ $n$-emission channel, we can write the isospin wavefunction of $RCN$ as a coupling of the two fragments,

\begin{equation}
\mid{T{RCN},T_{3RCN}}\rangle_n = \langle{T_{F1}T_{F2}T_{3F1}T_{3F2} \mid T_{RCN}T_{3RCN}}\rangle \mid{T_{F1},T_{3F1}}\rangle \mid{T_{F2},T_{3F2}}\rangle
\end{equation}
where $\langle T_{F1}T_{F2}T_{3F1}T_{3F2} \mid T_{RCN}T_{3RCN} \rangle$ represents the CG coefficients. The intensity of each fragment in the respective partition for a particular $n$-emission channel is, therefore, given by,
\begin{equation}
I_n = \langle{CGC}\rangle^2 = \langle{T_{F1}T_{F2}T_{3F1}T_{3F2} \mid T_{RCN}T_{3RCN}}\rangle^2
\end{equation}
The final yield of the fragment from all the $n$-emission channels may be obtained as,
\begin{equation}
I = \sum_{n} I_n \times w_n = \sum_{n} \langle{CGC}\rangle^2 \times w_n
\end{equation}
where $w_n$ is the normalized weight factor for $n^{th}$ $n$-emission channel taken from neutron multiplicity data in Fig. 5 of Bogachev et al.~\cite{bogachev}. We repeat the same procedure for all the lighter and heavier fragments of a partition. Then we normalize all the values with respect to the fragment having the maximum value. This will give us the relative yields of the fragments for that particular partition. We repeat this procedure for all the partitions.

\section{Results and discussions}

\begin{figure}
\centering
\includegraphics[width=0.60\textwidth]{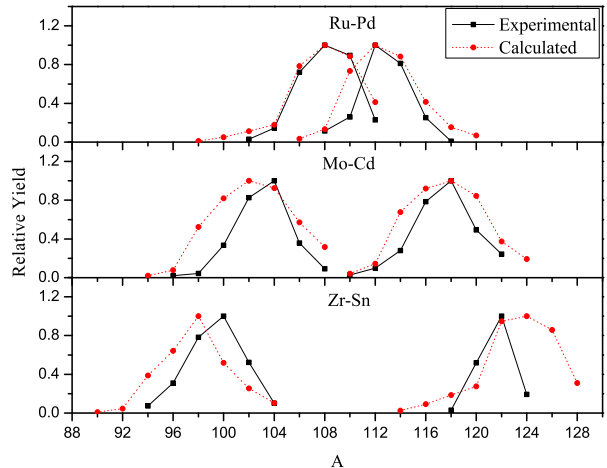}
\caption{Comparison of the calculated and experimental fission fragments yields vs. mass number A for the first three partitions only. Experimental data are taken from Bogachev et al.~\cite{bogachev}.}
\label{fig:boga}
\end{figure}

\begin{figure}
\centering
\includegraphics[width=0.60\textwidth]{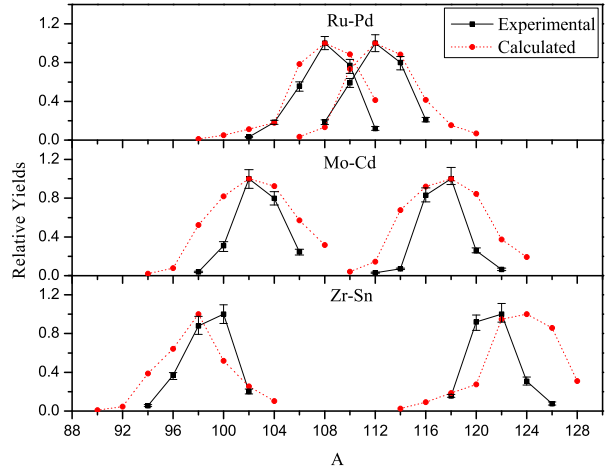}
\caption{Comparison of the calculated and experimental fission fragments yields vs. mass number A for the first three partitions only. Experimental data are taken from Banerjee et al.~\cite{banerjee} but the neutron multiplicity data has been taken from Bogachev et al.~\cite{bogachev}.}
\label{fig:bane}
\end{figure}

We present the results of our calculations for the relative yields of all the fragments emitted in $^{208}$Pb ($^{18}$O, f) for the first three partitions and then compare with the experimental data. We also normalize the experimental values as discussed in the above section for the calculated values. We compare our results with both the experimental data sets, Bogachev et al.~\cite{bogachev} and Banerjee et al.~\cite{banerjee}. However, we incorporate the same neutron multiplicity data taken from Bogachev et al. for both the calculations to include the weightage of various $n$-emission channels. Fig.~\ref{fig:boga} presents the comparison of relative yields of fission fragments  from our calculations and experimental data from Bogachev et al. A reasonably good agreement can be seen between the two, especially for the first two partitions  Ru-Pd, Mo-Cd. It is estimated that there may be 10-30\% error in the total relative yields in the experimental data~\cite{bogachev} which leads to a minimum 10\% error in the partition wise relative yields. Fig.~\ref{fig:bane} presents the comparison of the relative yields of fission fragments from our calculations and experimental data from Banerjee et al. In this case also, the calculations reproduce the experimental values very well for the first two partitions.

We note that the two experimental datasets do not match at several points. For example, in Mo-Cd partition data, peak is at $^{104}$Mo in Fig.~\ref{fig:boga} and at $^{102}$Mo in Fig.~\ref{fig:bane}. For the Zr-Sn partition, there is a shift of the peak in our calculated values from both the experimental data sets. One possible reason for this is due to the presence of $^{98}$Zr isomer with half-life $t_{1/2}$=1.7 $\mu$s. This lowers the experimental yield from what we expect from theory and we have not taken into account any corrections due to the presence of isomeric states in our calculations. Another possibility for the shift can be attributed to the presence of $A$ = 124 ($Z$ = 50) closed shell which may lead to the dip in the experimental yields at $A$ = 124 and its complementary fragment at $A$ = 98 as suggested by Danu et al.~\cite{danu}. But we do not consider any kind of shell effects in our calculations. Overall we can say that there is a reasonably good agreement between the calculated and experimental values which supports the concept of conservation of isospin in neutron-rich nuclei.

\section*{4. Conclusion}
We have calculated the relative yields of neutron-rich fission fragments emitted in $^{208}$Pb ($^{18}$O, f) and compared our calculated results with the experimental data from two sources~\cite{bogachev, banerjee}. For the calculations, we mainly use the theoretical argument given by Lane and Soper which suggests that the isospin impurity decreases with the neutron excess in heavy nuclei. Further, for the assignment of total isospin to the $RCN$ and the fission fragments, we invoke Kelson's conjectures. We have also incorporated the effect of neutron multiplicity by using the data from Bogachev et al. We assume that the relative yields of fission fragments are directly proportional to the square of $CGC$ obtained from the isospin part of the total wave function of $RCN$. The calculations are very simple based on isospin conservation. We have not included any kind of shell effects in our calculations which play a very important role in fission. Still, the calculations reproduce the relative yields of neutron-rich fission fragments quite reasonably. Thus, we conclude that the purity of isospin increases with neutron enrichment, which implies that the isospin becomes a reasonably good quantum number in neutron-rich systems. This may have important implications in many areas.

\section*{Acknowledgement}
We thank D. Robson for valuable comments and discussions. Support from Ministry of Human Resource Development (Government of India) to SG in the form of a fellowship is gratefully acknowledged. The authors also acknowledge the financial support in the form of travel grant from the Department of Science and Technology, Government of India and IIT Roorkee.

\end{document}